\documentclass[preprint,aps,prl,superscriptaddress,showpacs]{revtex4}
\usepackage{amsmath,amssymb,color,graphicx}

\begin{document}

\title{Synchronization and scaling properties of chaotic networks with multiple delays}
\author{Otti D'Huys}
\affiliation{Institute of Theoretical Physics, University of W\"urzburg, 97074 W\"urzburg,Germany}
\author{Steffen Zeeb}
\affiliation{Institute of Theoretical Physics, University of W\"urzburg, 97074 W\"urzburg,Germany}
\author{Thomas J\"ungling}
\affiliation{Institute of Theoretical Physics, University of W\"urzburg, 97074 W\"urzburg,Germany}
\affiliation{Instituto de F\'{i}sica Interdisciplinar y Sistemas Complejos, IFISC (UIB-CSIC), Campus Universitat de les Illes Balears, 07122 Palma de Mallorca, Spain}
\author{Serhiy Yanchuk}
\affiliation{Institute of Mathematics, Humboldt University of Berlin, 10099 Berlin, Germany}
\author{Wolfgang Kinzel}
\affiliation{Institute of Theoretical Physics, University of W\"urzburg, 97074 W\"urzburg,Germany}

\begin{abstract}
We study chaotic systems with multiple time delays that range over several orders of magnitude.  We show that the spectrum of Lyapunov exponents (LE) in such systems possesses a hierarchical structure, with different parts scaling with the different delays. This leads to different types of chaos, depending on the scaling of the maximal LE. Our results are relevant, in particular, for the synchronization properties of hierarchical networks (networks of networks) where the nodes of subnetworks are coupled with shorter delays and couplings between different subnetworks are realized with longer delay times. Units within a subnetwork can synchronize if the maximal exponent scales with the shorter delay, long range synchronization between different subnetworks is only possible if the maximal exponent scales with the long delay. The results are illustrated analytically for Bernoulli maps and numerically for tent maps.
\end{abstract}

\maketitle
Networks of nonlinear units with time-delayed interactions play an important role in various systems, such as coupled semiconductor lasers, predator/prey systems, traffic dynamics, communication networks, genetic circuits, or the brain \cite{Orosz09,Mat09,Chen2002,bookThomas,C,D}. Delay times may induce high-dimensional chaotic dynamics \cite{Farmer1982,Lepri1994,Giacomelli1996}, as for example in semiconductor lasers with delayed feedback \cite{AHL98}. A particularly interesting phenomenon in this context is the zero lag synchronization of chaotic units, despite the long interaction delays \cite{Fischer06,Klein2006,MUR10,RAV11,PEI07}. Chaos synchronization finds applications in encrypted communication \cite{arg05.1,Uchida2012}.

Chaos in the network is quantified by the spectrum of Lyapunov exponents (LEs), which measures the sensitivity to initial conditions. For a system with one long delay, chaos can be characterized by the scaling of the maximal LE with increasing delay time $\tau$. In the region of strong chaos, it approaches a positive constant value whereas for weak chaos the maximal LE decreases as $1/\tau$. These scaling properties have consequences for chaos synchronization: Networks with strong chaos cannot synchronize completely for long delays, whereas for weak chaos synchronization is possible depending on the value of the maximal LE and the topology of the network \cite{PRLValentin,Heiligenthal2011}.

Strong and weak chaos have been demonstrated in networks with a single time delay, up to now. However, realistic systems may have different transmission delays for the coupling signals. In a network with a distribution of delays complex behavior is expected to be suppressed \cite{Marti2006,Masoller2005,ata03}. However, if the network has different delay times with special integer ratios, one finds resonances which can either stabilize or rule out chaos synchronization \cite{Englert10}, depending on the ratio.

In this Letter we study networks of nonlinear units coupled by multiple delay times which differ by several orders of magnitude. A typical example is a network of networks, with a connection delay $\tau_{1}$ between the nodes within a sub-network and a much longer connection delay $\tau_{2}$ between the different sub-networks. We explain the scaling of the full spectrum of LEs with increasing delay times and extend the concepts of strong and weak chaos to multiple delay systems. Finally, we relate the synchronization properties of a hierarchical network to the scaling behavior of the LEs.

\paragraph{Spectrum of Lyapunov exponents of a Bernoulli map.}

Strong and weak chaos have been found both for time-continuous and discrete sytems with delay \cite{Lepri1994,ENG11}. Since the main results are valid in both cases, we perform our calculations for iterated maps. For networks of Bernoulli maps, even analytic results can be derived. We start with a single chaotic map with $N$ different feedback delays $\tau_1,\hdots,\tau_N$, described by
\begin{eqnarray}
x_{t+1} & = & (1-\epsilon)f(x_{t})+\epsilon{\displaystyle \sum\limits _{k=1}^{N}\kappa_{k}f(x_{t-\tau_{k}})\,,}\label{eq:eqsys}
\end{eqnarray}
with $x\in\mathbb{R}$. We consider here delays with different orders of magnitudes $1\ll\tau_{1}\ll\hdots\ll\tau_{N}$. The spectrum of LEs $\Lambda=\{\lambda_1,\hdots,\lambda_N\}$, which describes the evolution of a perturbation $\delta x_{t}$ along a chaotic trajectory $s_{t}$, is calculated using the linearized equation 
\begin{equation}
\delta x_{t+1}=(1-\epsilon)f'(s_{t})\delta x_{t}+\epsilon{\displaystyle \sum\limits _{k=1}^{N}\kappa_{k}f'(s_{t-\tau_{k}})\delta x_{t-\tau_{k}}\,.}\label{eq:lin}
\end{equation}
The coefficients $f'(s_{t})$ in general depend on time, since the trajectory $s_{t}$ is time dependent. However, for a Bernoulli map, given by 
\begin{equation}
f(x)=ax\mod1\,
\end{equation}
with $|a|>1$, we have constant coefficients $f'(x)=a$. The linearized equation \eqref{eq:lin} reduces then to a polynomial equation for the characteristic multipliers $z$, which characterize the growth of a perturbation $\delta x_{t}=\delta x_{0}z^{t}$ : 
\begin{equation}
z=(1-\epsilon)f'+\epsilon f'{\displaystyle \sum\limits _{k=1}^{N}\kappa_{k}z^{-\tau_{k}}\,.}\label{eq:eqchar}
\end{equation}
The Lyapunov exponents $\lambda\in\Lambda$ are given by $\lambda=\ln|z|$. The calculation of the LEs can thus be performed in the same way as the stability calculation of a steady state. 
For a single delay system, it is known that such an equation can have two different types of unstable (with $\left|z\right|>1$) solutions \cite{Lepri1994,WOL06}: strongly unstable and weakly unstable. The strongly unstable root is approximated by the delay-independent term $z_0\approx(1-\epsilon)f'$, provided $|(1-\epsilon)f'|>1$, and does not depend on the delays to the leading order. Also the multiple-delay system can have a strongly unstable multiplier.

In analogy with the single delay system, we assume a scaling behavior for the next group of multipliers $\ln z=i\omega+\gamma_{1}/\tau_{1}$, with $\gamma_{1}>0$. In leading order, by neglecting all terms of order $1/\tau_{1}$, $e^{-\gamma_1\tau_2/\tau_1}$, and smaller, the characteristic polynomial \eqref{eq:eqchar} reads 
\begin{equation}
e^{i\omega}=(1-\epsilon)f'+\epsilon\kappa_{1}f'e^{-i\omega\tau_{1}-\gamma_{1}}\,.
\end{equation}
This equation allows to compute a curve $\gamma_{1}(\omega)=\ln|\epsilon\kappa_{1}f'|-\ln|e^{i\omega}-(1-\epsilon)f'|$, on which the roots are located. The imaginary parts of these roots differ by approximately $\Delta\omega\approx2\pi/\tau_{1}$, the number of roots thus increases linearly with $\tau_{1}$. In the following we call the unstable subset of these LEs the $\tau_{1}$-spectrum $\Lambda_{1}$. The $\tau_{1}$-spectrum corresponds to the pseudo-continuous spectrum for steady states of single-delay systems \cite{WOL06}.

For the third group of unstable multipliers we assume a scaling with the second delay $\ln z=i\omega+\gamma_{2}/\tau_{2}$, with $\gamma_{2}>0$. In leading order, we find a characteristic polynomial 
\begin{equation}
e^{i\omega}=(1-\epsilon)f'+\epsilon\kappa_{1}f'e^{-i\omega\tau_{1}}+\epsilon\kappa_{2}f'e^{-i\omega\tau_{2}+\gamma_{2}}.
\end{equation}
In an analogous way, the unstable roots (if they exist) are approximated by a curve $\gamma_{2}(\omega)=\ln|\epsilon\kappa_{2}f'|-\ln|e^{i\omega}-(1-\epsilon)f'-e^{-i\omega\tau_{1}}\epsilon\kappa_{1}f'|$. The corresponding LEs form a $\tau_{2}$-spectrum $\Lambda_{2}$, which scales inversely with the second delay $\tau_{2}$. The number of exponents in this spectrum scales linearly with $\tau_{2}$.

\begin{figure}\begin{minipage}{4cm}
\includegraphics[height=2.8cm]{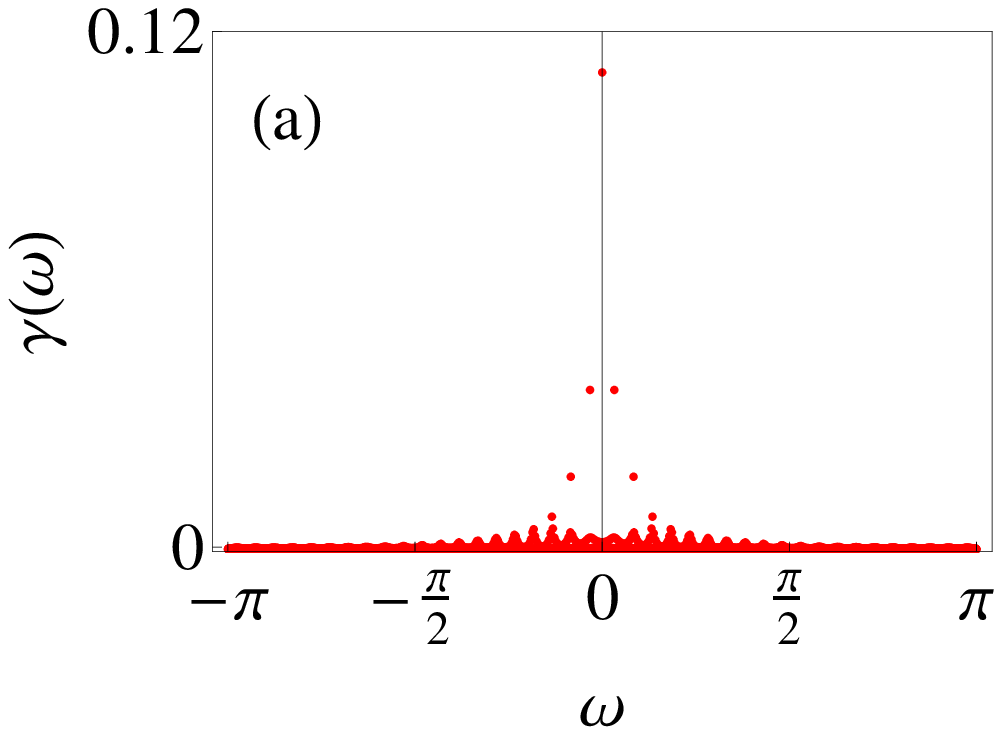}
\includegraphics[height=2.8cm]{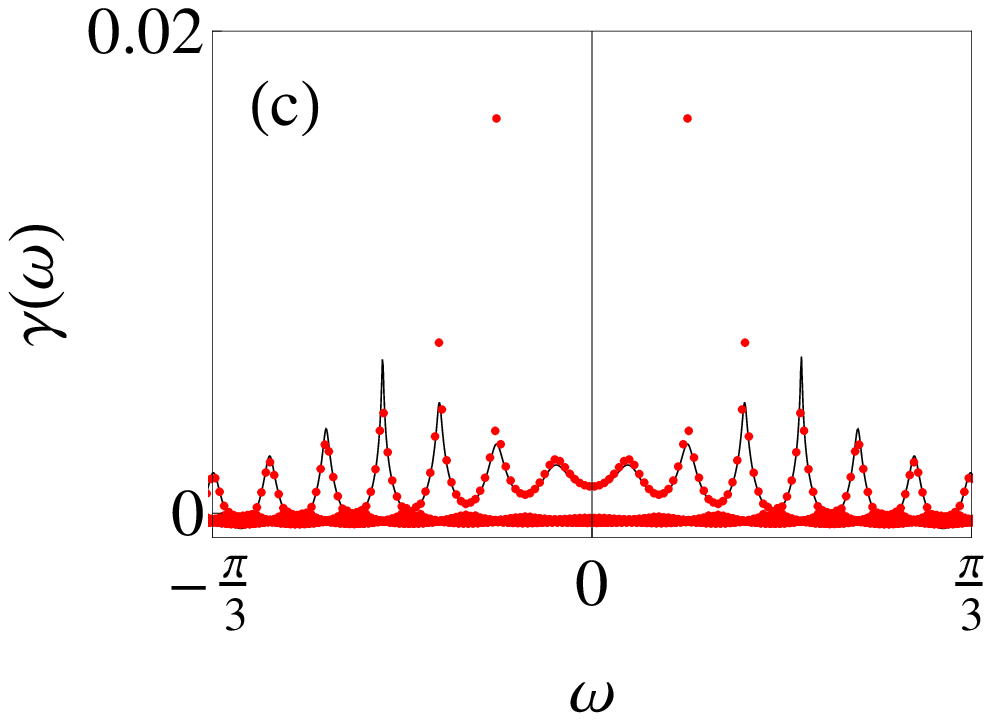}
\end{minipage}
\begin{minipage}{4cm}
\begin{flushright}\includegraphics[height=2.8cm]{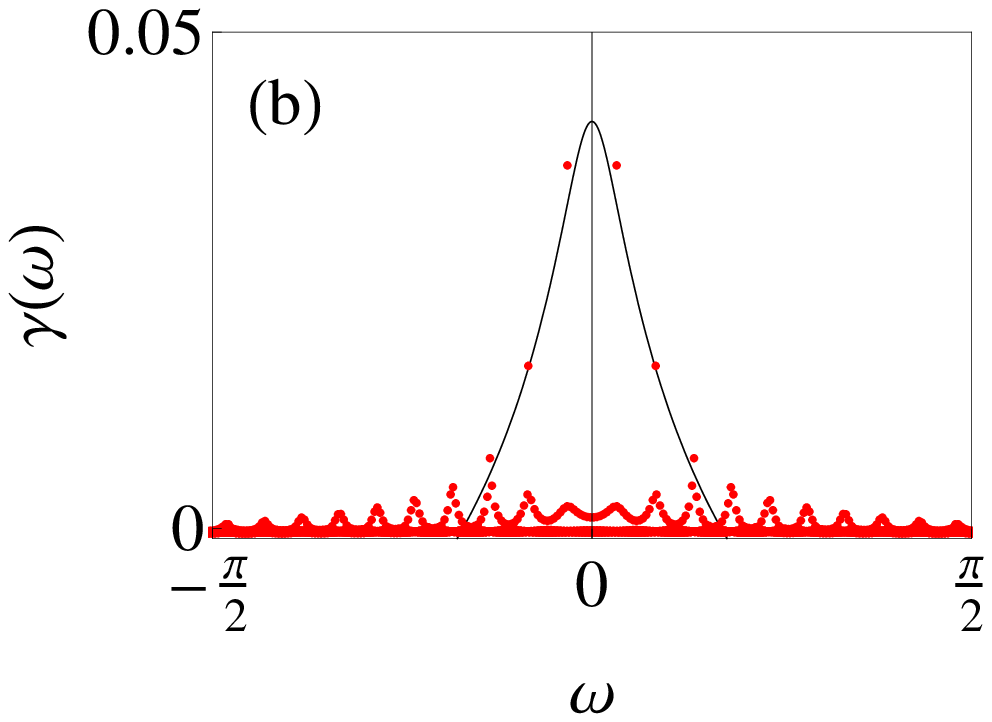}\newline\includegraphics[height=2.8cm]{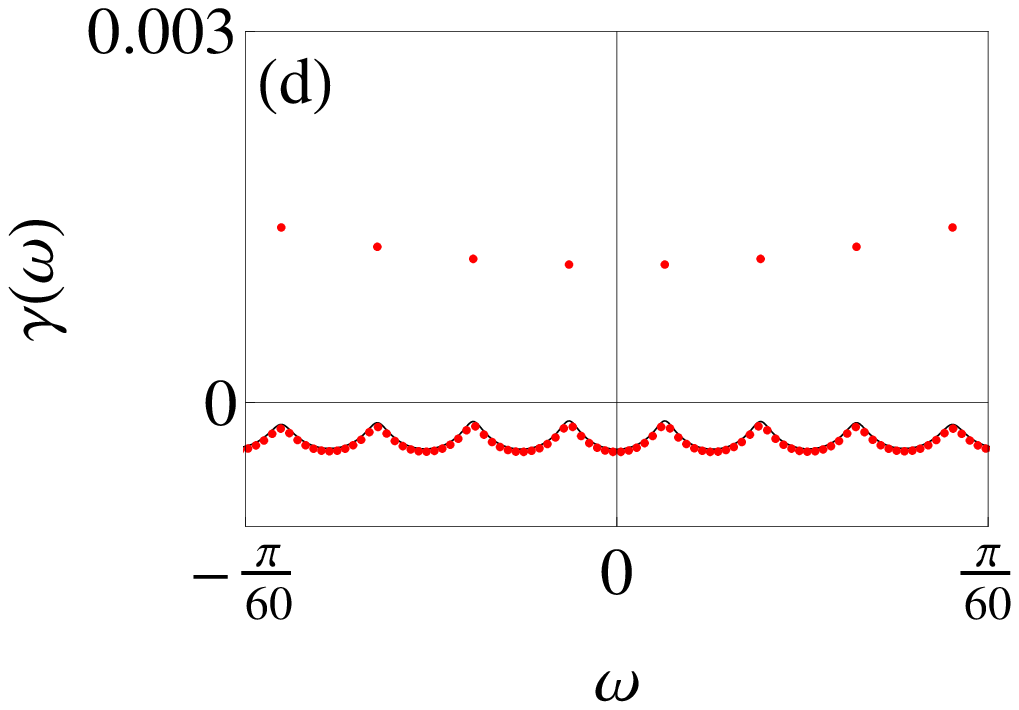}\end{flushright}\end{minipage}
\caption{(Color online) The spectrum of LEs of a Bernoulli map (Eq. \eqref{eq:eqchar}) (red dots) subject to three feedback delays. The different panels are zooms.  The full black lines are the analytic approximations for long delays $\gamma_{1}(\omega)/\tau_{1}$ (panel (b)), $\gamma_{2}(\omega)/\tau_{2}$ (panel (c)) and $\gamma_{3}(\omega)/\tau_{3}$ (panel (d)). Parameters are $a=3$, $\epsilon=0.63$, $\kappa_{1}=0.3$, $\kappa_{2}=0.6$, $\kappa_{3}=0.1$, $\tau_{1}=40$, $\tau_{2}=500$ and $\tau_{3}=6000$.}
\label{fig:spec} 
\end{figure}

It is possible to calculate unstable spectra $\Lambda_{k}$ related to each of the delays $\tau_{k}$; only the $\tau_{N}$-spectrum, related to the largest delay present in the system, can have a stable part. Figure \ref{fig:spec} shows the spectrum of LEs for a Bernoulli map, obtained by solving Eq. \eqref{eq:eqchar} numerically, and the analytical long delay approximations $\gamma_{k}(\omega)$. Although the different time scales are not so far apart, the analytical curves $\gamma_{k}(\omega)$  provide a good approximation for the LEs. We can clearly distinguish one strongly unstable multiplier, and the weakly unstable spectra with their respective delay-scaling. 

A similar hierarchy of eigenvalues for steady states of time-continuous delay differential equations with multiple delays can be shown applying the same arguments to the corresponding transcendental equation.

\paragraph{General case.}

The slope of a chaotic map in general depends on the trajectory, so that the linearization (Eq. \eqref{eq:lin}) is time-dependent. The Lyapunov spectrum is then evaluated numerically using a Gram-Schmidt orthogonalization procedure according to Farmer \cite{FAR82}. We find that the properties which we derived for the time-independent case are preserved in the presence of fluctuations, so that the complete Lyapunov spectrum $\Lambda$ is composition of $\tau_{k}$-spectra. The $\tau_{k}$-spectrum is obtained numerically by integrating the evolution of an auxiliary perturbation variable $\delta x_{t}^{k}$, for which the delay terms $\tau_{k+1}$ to $\tau_{N}$ are removed 
\begin{equation}
\begin{split}\delta x_{t+1}^{0} & =(1-\epsilon)f'(s_{t})\delta x_{t}^{0}\\
 & \Rightarrow\Lambda_{0}=\{\lambda_{0}\}\\
\delta x_{t+1}^{1} & =(1-\epsilon)f'(s_{t})\delta x_{t}^{1}+\epsilon\kappa_{1}f'(s_{t-\tau_{1}})\delta x_{t-\tau_{1}}^{1}\\
 & \Rightarrow\Lambda_{1}=\{\lambda_{1,\max},\dots,\lambda_{1,n}\}\setminus\Lambda_{0}\\
 & \vdots
\end{split}
\label{eq:numspecs}
\end{equation}
The first exponent $\lambda_0$ is called the sub-LE. If a partial spectrum $\Lambda_{k}$ contains positive exponents, they can be said to 'survive' the introduction of further time scale separated delay terms, because the contribution of the additional terms becomes exponentially small for the corresponding Lyapunov modes. We exclude these exponents from the definition of the succeeding spectra, so that the partial spectra $\Lambda_{k}$ do not overlap, and each spectrum $\Lambda_k$ scales only inversely with $\tau_k$.

\begin{figure}
\begin{minipage}{4cm}
\begin{flushright}\includegraphics[width=0.95\columnwidth]{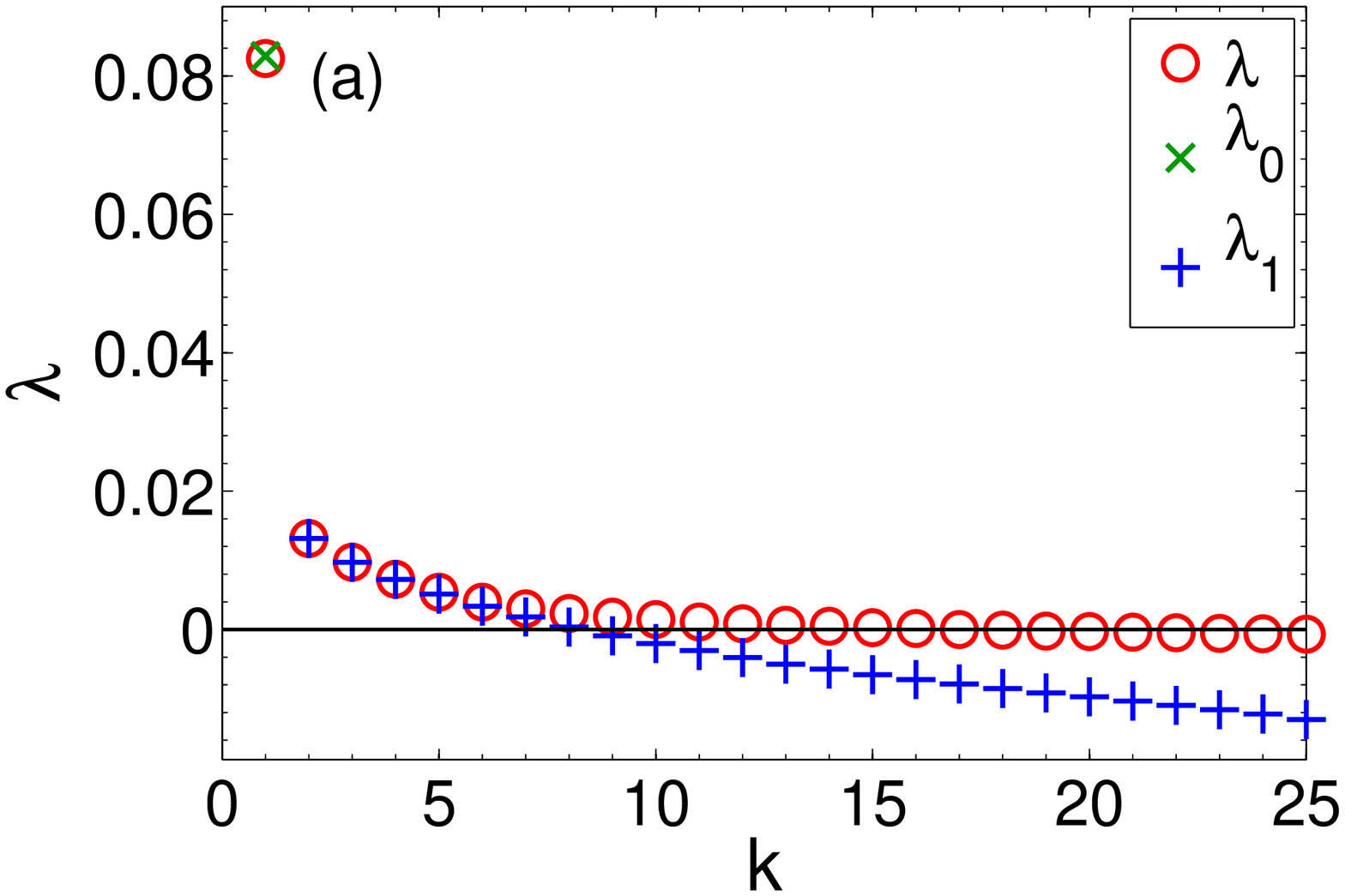}\end{flushright}
\includegraphics[width=0.96\columnwidth]{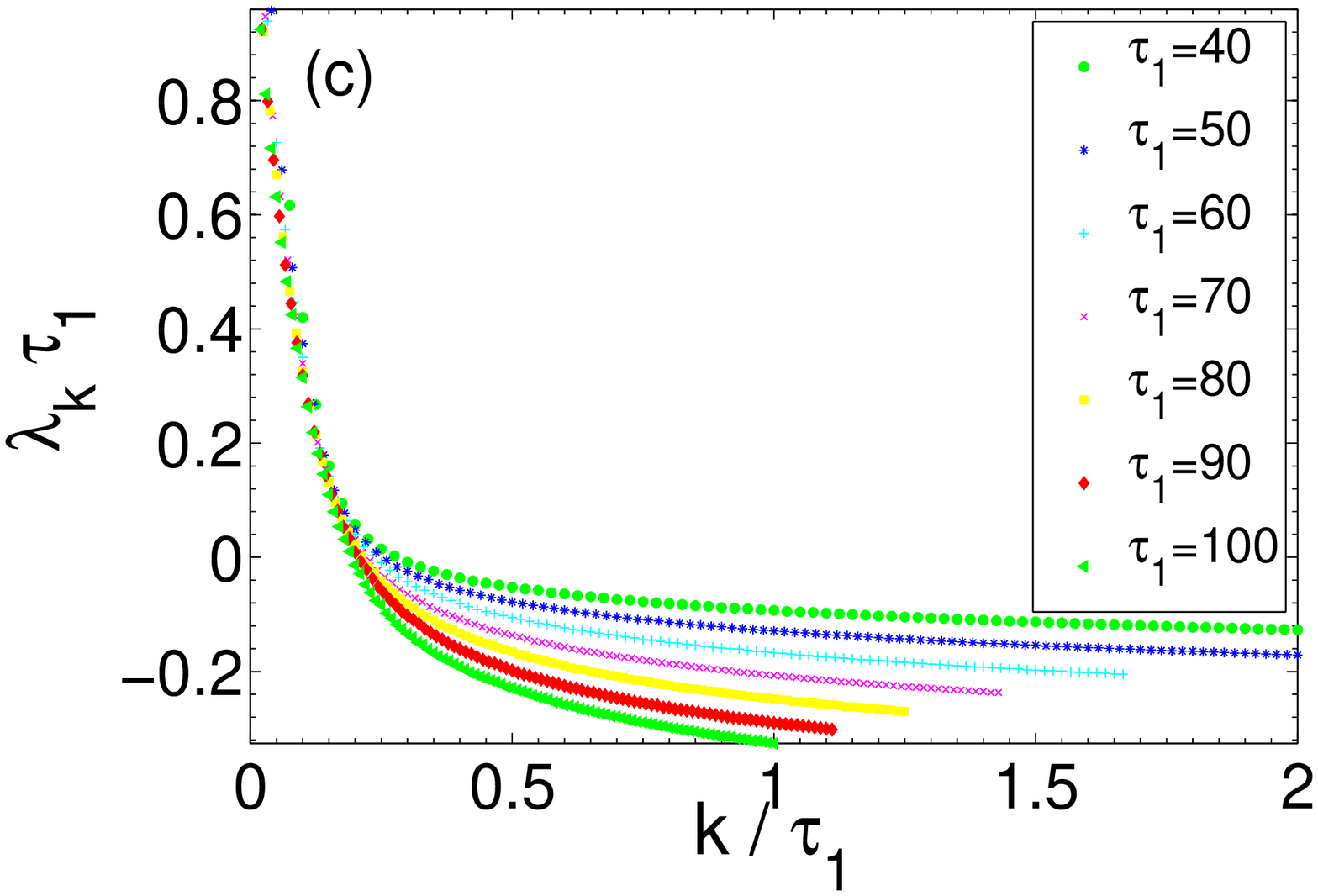}
\end{minipage}
\begin{minipage}{4cm}
\includegraphics[width=0.95\columnwidth]{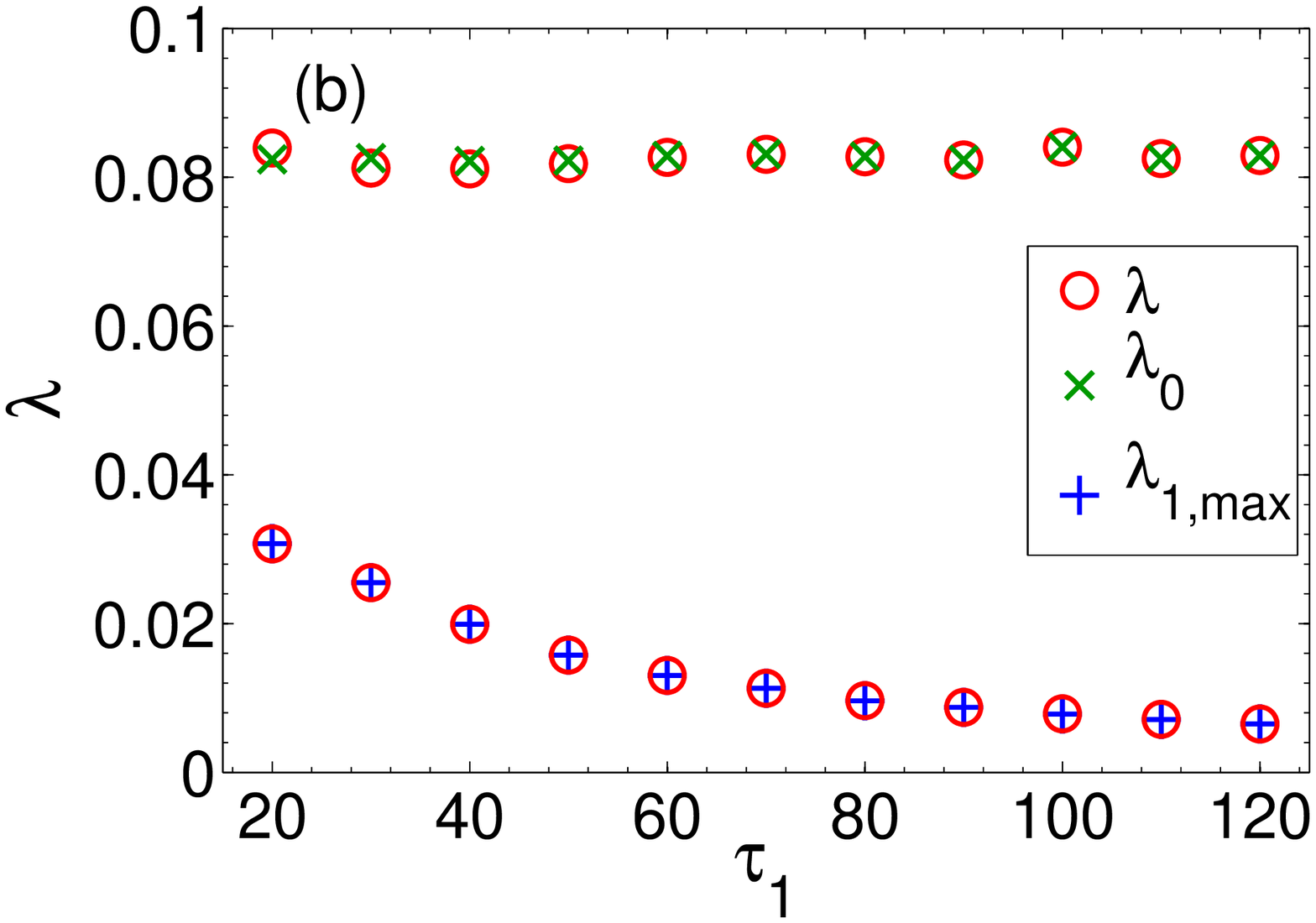}
\begin{flushright}\includegraphics[width=0.96\columnwidth]{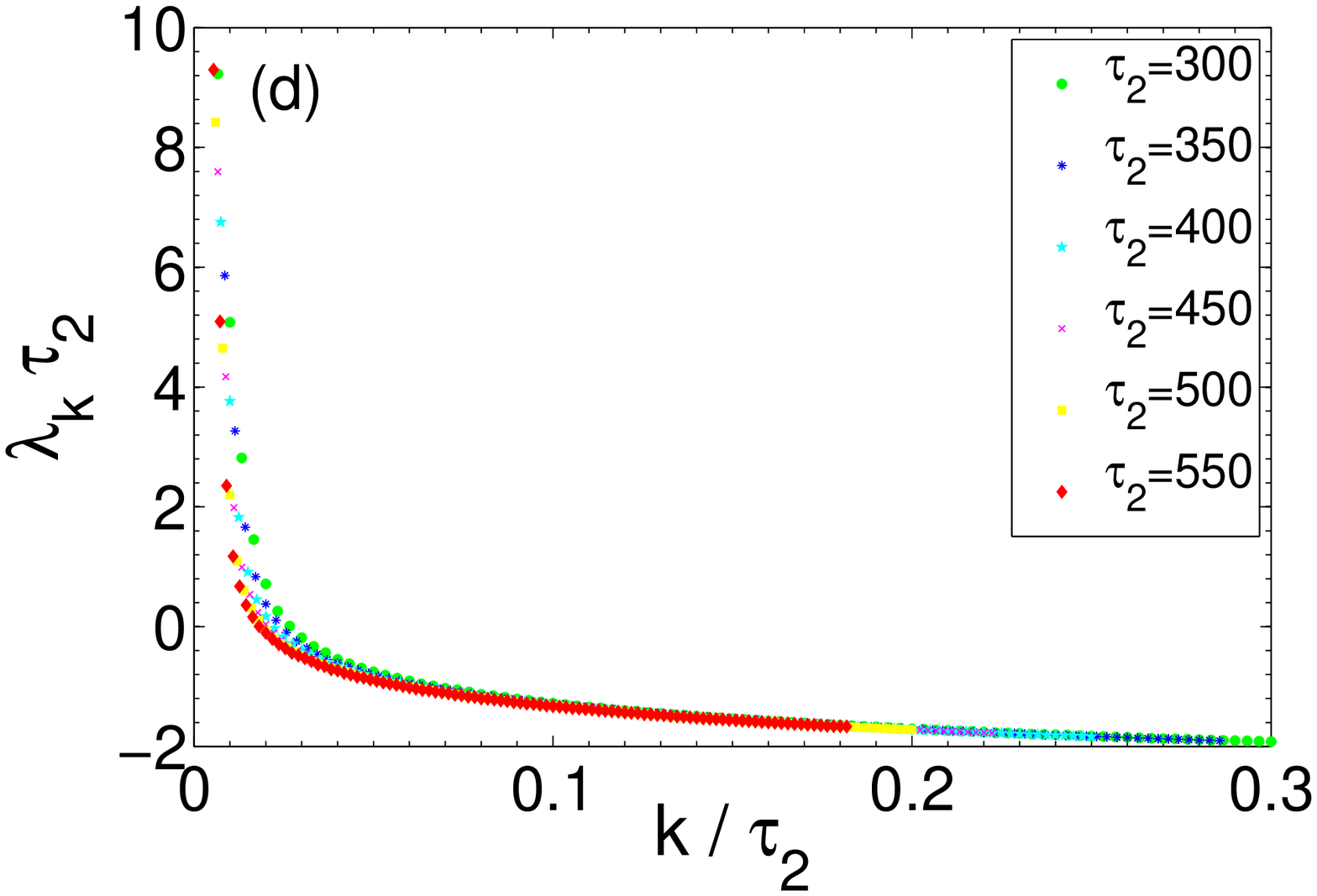}\end{flushright}
\end{minipage}
\caption{(Color online) Panel (a) shows $\lambda_{0}$ and the first 15 exponents of the full and the partial Lyapunov spectra, $\lambda$ and $\lambda_{1}$, for a tent map with two feedbacks $\tau_{1}=60$ and $\tau_{2}=500$ in the strong chaos regime. Panel (b) shows the sub-exponent $\lambda_{0}$ (green 'x' symbols),  two maximal LE of the full spectrum $\lambda$ (red circles), and the maximal exponent of the $\tau_{1}$-spectrum $\lambda_{1,\max}$ (blue crosses) in the same setup for different delays $\tau_{1}$. In panel (c) and (d) the LE $\lambda_{k}$, $k=2,\dots,100$ for different values of the feedback delays are shown. Panel (c) shows the scaling of the positive LEs with $\tau_{1}$ for fixed $\tau_{2}=500$. Panel (d) shows the $\tau_{2}$-scaling of the smaller exponents for fixed $\tau_{1}=30$. Other parameters are $a=0.4$, $\epsilon=0.4$, $\kappa_{1}=0.8$, and $\kappa_{2}=0.2$}
\label{fig:LyapunovSpectra} 
\end{figure}

We demonstrate the composition of the LE spectrum into different partial spectra for a tent map, 
\begin{equation}
f(x)=\left\{ \begin{array}{ll}
\frac{1}{a}\, x & \mathrm{for}\:0\leq x<a\\
\frac{1}{1-a}\,(1-x) & \mathrm{for}\: a\leq x\leq1
\end{array}\right.\,,\label{eq:tent}
\end{equation}
subject to two different delayed feedbacks. Figure \ref{fig:LyapunovSpectra}(a,b) compares the sub-LE $\lambda_{0}$, the partial spectrum $\Lambda_{1}$ and the total spectrum $\Lambda$. We find that the maximal LE $\lambda_{\max}\in\Lambda$ is well approximated by the sub-LE $\lambda_{0}$ for a long enough delay $\tau_{1}$. Since $\lambda_{0}$ depends on the trajectory $s_{t}$, it also depends indirectly on the delays $\tau_{k}$. Nevertheless, this dependence is negligible, as has also been reported for the Lang-Kobayashi model representing a single delayed feedback system \cite{Heiligenthal2011,PRLValentin}. The next exponents $\lambda_{2},\dots,\lambda_{7}$ of the full spectrum are approximated by the $\tau_{1}-$spectrum $\Lambda_{1}$; the full spectrum $\Lambda$ deviates from $\Lambda_{1}$ as the latter becomes negative. The second LE $\lambda_{2}$, which coincides with $\lambda_{1,\max}$, decreases with $\tau_{1}$.

The different scaling behaviors of the LE spectrum are depicted in Fig.~\ref{fig:LyapunovSpectra}(c,d). The partial spectrum $\Lambda_1$ as a whole (and thus the upper part of the full spectrum) scales inversely with the delay $\tau_1$; the number of exponents in the partial spectrum however scales linearly with $\tau_1$. These two effects are demonstrated by plotting $\lambda_{k}\tau_{1}$ vs.~$k/\tau_{1}$  (with $k$ the ranking of the exponent): For different delays $\tau_{1}$ the spectrum converges to a curve for all exponents from the $\tau_{1}$-spectrum.  The curves diverge for smaller exponents $k/\tau_{1}\gtrsim0.2$, as these scale with the largest delay $\tau_{2}$. Similar, for varying $\tau_{2}$ the spectrum converges to a curve for all exponents $\lambda_{k}$ from $\Lambda_{2}$ when plotting $\lambda_{k}\tau_{2}$ vs.~$k/\tau_{2}$.

\paragraph{$\tau_{k}$-chaos.}

Apart from a hierarchical Lyapunov spectrum as described above, the different time scales can also manifest in the maximal LE $\lambda_{\max}$, which can be considered as the most important quantity describing a chaotic system. If $\lambda_{0}>0$ we speak of strong chaos; the maximal LE $\lambda_{\max}\approx\lambda_{0}$ and does not vary with any of the delays. In the weakly chaotic regimes $\lambda_{0}<0$ holds. If $\lambda_{1,\max}>0$ we speak of $\tau_{1}$-chaos and $\lambda_{\max}\propto1/\tau_{1}$. If $\lambda_{1,\max}<0$ the second delay dominates and $\lambda_{\max}\propto 1/\tau_2$. Consequently we define $\tau_{k}$-chaos as the scaling of $\lambda_{\max}$ with $1/\tau_{k}$.

\begin{figure}
\includegraphics[height=3.15 cm]{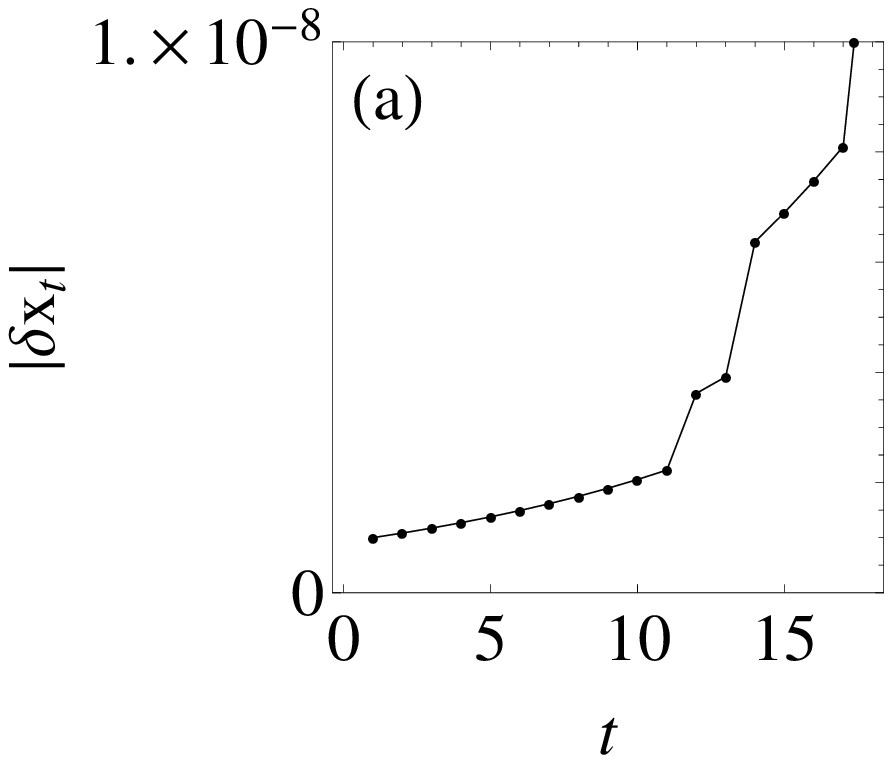}
\includegraphics[height=3 cm]{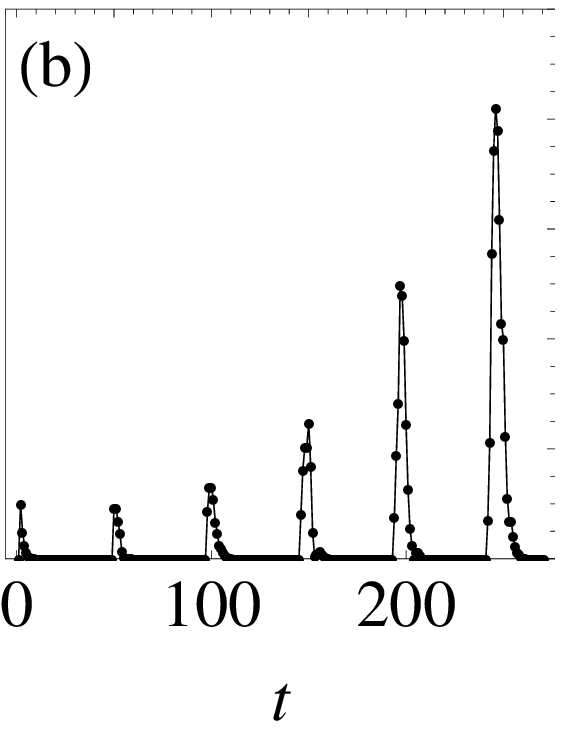}
\includegraphics[height=3 cm]{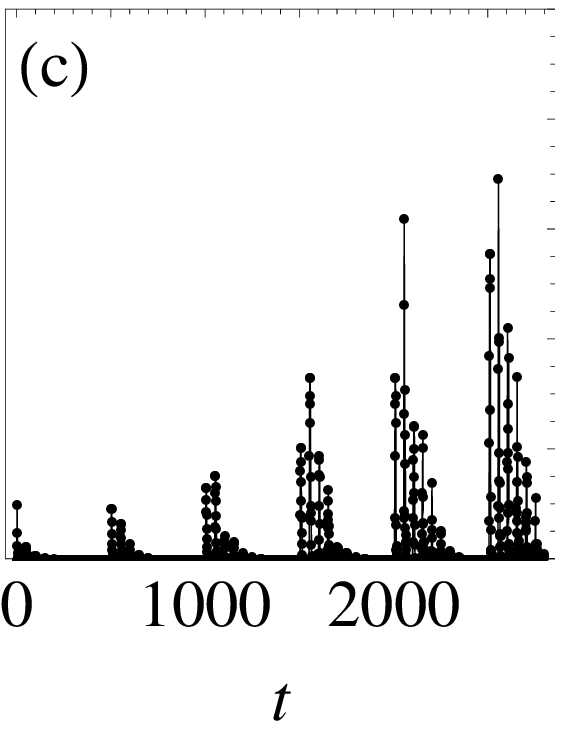}
\caption{Evolution of a small perturbation for a tent map with two delayed feedbacks, for (a) strong chaos ($\epsilon=0.35$, $\kappa=0.8$), (b) $\tau_{1}$-chaos ($\epsilon=0.7$, $\kappa=0.8$) and (c) $\tau_{2}$-chaos ($\epsilon=0.7$, $\kappa=0.2$). Other parameters are $\tau_{1}=47$, $\tau_{2}=500$ and $a=0.4$}
\label{fig:perturb} 
\end{figure}

The difference between strong, $\tau_{1}$-, $\tau_{2}$-,$\dots$, $\tau_{k}$-chaotic dynamics can be directly observed in the evolution of a small perturbation.  Fig. \ref{fig:perturb} shows the difference $\delta x_t$ between trajectories of two identical chaotic tent maps, initialized identically except for a point-like perturbation at $t=0$. The instantaneous evolution of this perturbation is governed by $\lambda_{0}$. In the strongly chaotic regime, it thus increases exponentially, as shown in Fig. \ref{fig:perturb}(a). In the weakly chaotic regimes the perturbation decays first, to reappear and decay at $t\approx\tau_{1}$. The perturbation evolves over the consecutive $\tau_{1}$-intervals according to $\lambda_{1,\max}$. In the $\tau_{1}$-chaotic regime a perturbation thus spreads on the time scale of $\tau_{1}$ (illustrated in Fig. \ref{fig:perturb}(b)). In the $\tau_{2}$-chaotic regime it decays over the $\tau_{1}$-intervals, but it is magnified over the $\tau_{2}$-intervals. The behavior is shown in Fig. \ref{fig:perturb}(c). 

\paragraph{Synchronization.}

The sub-LE $\lambda_{0}$ and the $\tau_{1}$-spectrum $\Lambda_{1}$ appear naturally in the context of synchronization in a network of networks. In this case the interaction delays $\tau_{1}$ within a subnetwork are much shorter than the connection delays $\tau_{2}$ between the different subnetworks. We illustrate this with a simple hierarchical network of dynamical units, described by 
\begin{eqnarray}
x_{t+1}^{jm} & = & (1-\epsilon)f(x_{t}^{jm})+\epsilon\kappa\sum_{l}A_{jl}^{(m)}f(x_{t-\tau_{1}}^{lm})\nonumber \\
 &  & +\epsilon(1-\kappa)\sum_{sk}B_{jk}^{(ms)}C_{ms}f(x_{t-\tau_{2}}^{ks})\,.\label{eq:network}
\end{eqnarray}
The coupling topology within the $m$-th subnetwork is described by the matrix $A^{(m)}$, the matrix $C$ describes the coupling architecture of the networks, and the matrix $B^{(ms)}$ models the coupling between the $m$-th and the $s$-th subnetwork. We assume that all the elements receive the same amount of input, both from within and from outside the subnetwork (i.e. all the matrices have a row sum equal to $1$), so that all the nodes can synchronize identically. Moreover, we consider identical rows for the $B$-matrices, so that all the elements of a subnetwork receive the same input from outside.

To determine the stability of a (cluster) synchronized state $s_{t}^{m}$, the network model Eq. \eqref{eq:network} is linearized along the corresponding synchronization manifold. Evaluating this linear system along the transverse eigendirections yields
\begin{equation}
\delta x_{t+1}=(1-\epsilon)f'(s_{t}^{m})\delta x_{t}+\sigma_{A}^{(m)}\kappa\epsilon f'(s_{t-\tau_{1}}^{m})\delta x_{t-\tau_{1}}\,,\label{eq:msf-subnetwerk}
\end{equation}
with $\sigma_{A}^{(m)}$ the transverse eigenvalues of the connection matrix $A^{(m)}$. Integration of Eq. \eqref{eq:msf-subnetwerk} reveals the master stability function $\lambda(\sigma_A)$ \cite{PEC90}. The explicit dependence on the long delay connections vanishes, as all the nodes receive the same external input.  Hence, Eq.\eqref{eq:msf-subnetwerk} and the corresponding master stability function take the same form as for a network with a single delay \cite{choe10,Heiligenthal2011}. The difference lies within the dynamics of the synchronized state $s_{t}^{m}$, which now also depends on the elements outside the subnetwork and the connection delay $\tau_{2}$. In the strongly chaotic regime, the largest transverse LE is approximately $\lambda_{0}>0$, so that the elements cannot synchronize. In the $\tau_{1}$-chaotic regime, identical or cluster synchronization in the $m$-th subnetwork is stable if $|\sigma_{A}^{(m)}|\leq e^{-\lambda_{\max}\tau_{1}}$ \cite{Heiligenthal2011}; the synchronization pattern in the subnetwork thus depends on the coupling topology $A^{(m)}$. For $\tau_{2}$-chaotic behavior all the nodes of a subnetwork synchronize completely irrespective of the coupling architecture: the elements of the subnetwork show consistent behavior with respect to the common input from the rest of the network.

The stability of identical synchronization of the full network is then governed by the following equation
\begin{eqnarray}
\delta x_{t+1} & = & (1-\epsilon)f'(s_{t})\delta x_{t}+\epsilon\kappa f'(s_{t-\tau_{1}})\delta x_{t-\tau_{1}}\nonumber \\
 &  & +\sigma_{C}\epsilon(1-\kappa)f'(s_{t-\tau_{2}})\delta x_{t-\tau_{2}}\,,
\end{eqnarray}
with $\sigma_{C}$ the transverse eigenvalues of the intra-network coupling matrix $C$. Synchronization between subnetworks is only possible in the $\tau_{2}$-chaotic regime in the limit of long delays, on the condition that $|\sigma_{C}|\leq e^{-\lambda_{\max}\tau_{2}}$ holds. 

\begin{figure}
\begin{minipage}[c]{3 cm}%
 \includegraphics[width=1\columnwidth]{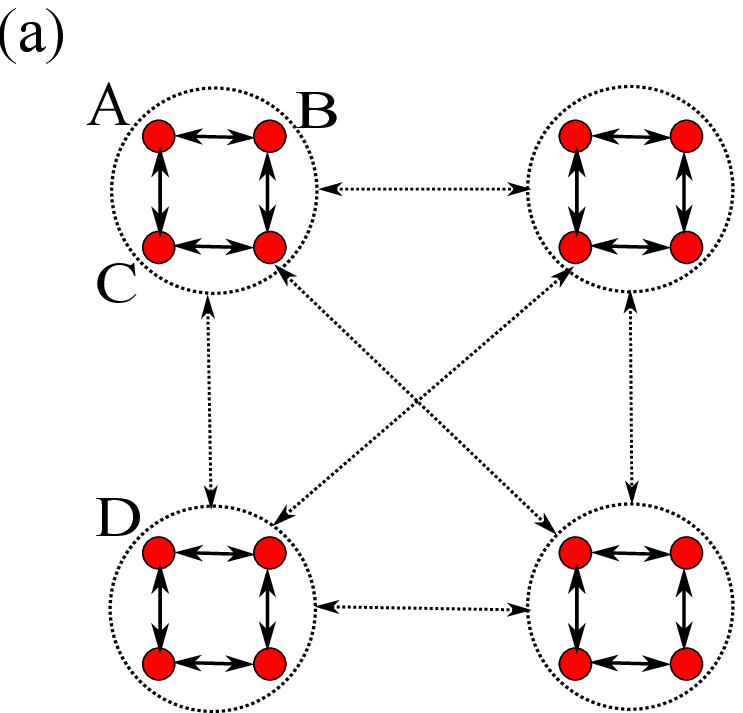} %
\end{minipage}%
\begin{minipage}{3.5cm}%
\begin{flushright}
 \includegraphics[width=1\columnwidth]{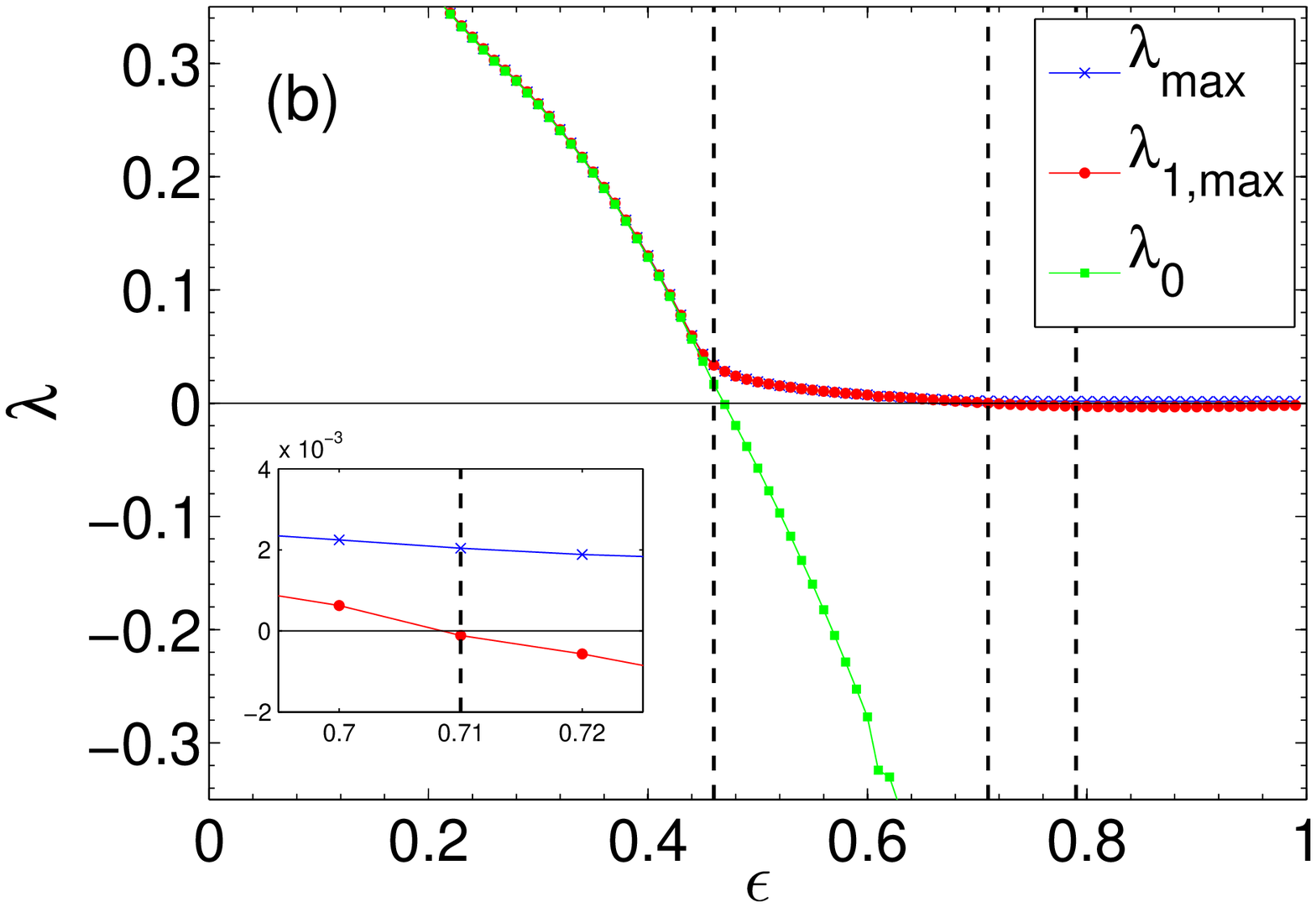} \includegraphics[width=0.96\columnwidth]{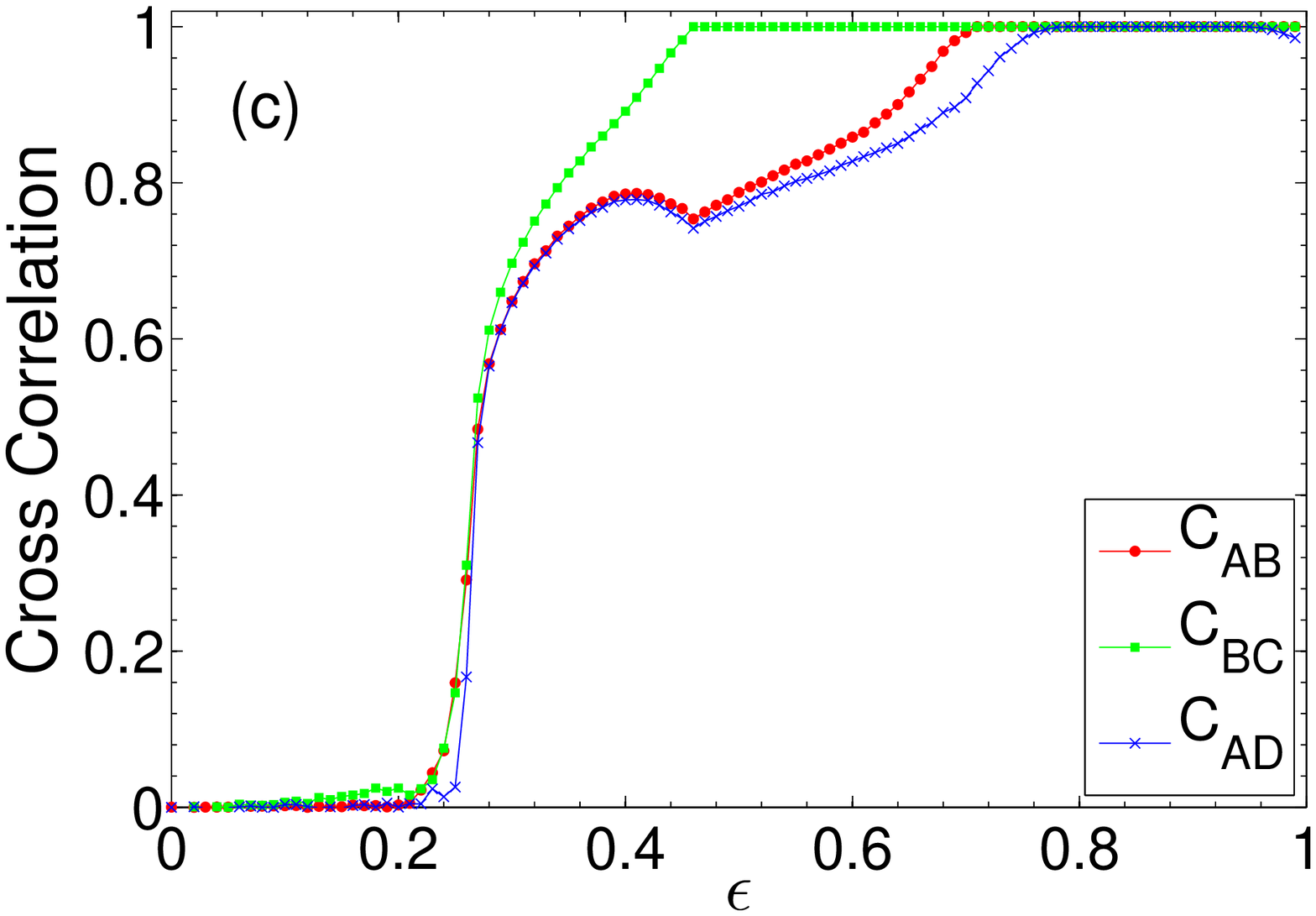}\end{flushright} %
\end{minipage}\caption{(Color online) Panel (a) shows a sketch of the network topology. In panel (b) the sub-LE $\lambda_{0}$ and $\lambda_{1,\max}$ and the maximal LE $\lambda_{\max}$ are plotted for a hierarchical network of tent maps (described by Eqs. \eqref{eq:network} and \eqref{eq:tent}) as a function of the total coupling strength $\epsilon$. The dashed lines indicate the synchronization transitions. Panel (c) shows the crosscorrelations between nodes $A$ and $B$ (red dots), $B$ and $C$ (green squares) and $A$ and $D$ (blue crosses) for varying total coupling strength $\epsilon$. Parameters are $\alpha=0.4$, $\kappa=0.5$, $\tau_{1}=50$, and $\tau_{2}=500$.}
\label{fig:networkofnetworks} 
\end{figure}

As an example we consider a network of four globally coupled subnetworks, coupled through their mean fields (Fig. \ref{fig:networkofnetworks}(a)). The subnetworks are bidirectional rings of four elements. As individual dynamics we choose again tent maps (Eq. \eqref{eq:tent}). We increase the total coupling strength $\epsilon$, such that the system undergoes a transition from strong to $\tau_{1}$- to $\tau_{2}$-chaos. Figure \ref{fig:networkofnetworks} shows the crosscorrelations between several different network elements as function of $\epsilon$, together with the different sub-LEs $\lambda_{0}$ and $\lambda_{1,\max}$. The sublattice synchronization between the diagonal elements $B$ and $C$ in a subnetwork is governed by Eq. \eqref{eq:msf-subnetwerk}, with a transverse eigenvalue $\sigma_{A}=0$. The nodes $B$ and $C$ thus synchronize when $\lambda_{0}<0$. The small difference between the two transition points is caused by numerical inaccuracy. The synchronization between all elements in a subnetwork is governed by the same equation (Eq. \eqref{eq:msf-subnetwerk}), but the transverse eigenvalue with maximal magnitude is given by $\sigma_{A}=-1$. Consequently, the nodes $A$ and $B$ synchronize when $\lambda_{1}<0$. The inset in Fig. \ref{fig:networkofnetworks} shows an exact agreement between these two points. We find that the whole network, and thus the nodes $A$ and $D$, synchronizes for a slightly higher coupling strength.

In conclusion, we showed that a hierarchy of time scales emerges in systems with several delays. These time scales can be characterized by the different components in the spectrum of LEs. Depending on the leading components of the spectrum, one can distinguish between strong chaos or $\tau_{k}$-chaos. In the $\tau_{k}$-chaotic regime, small perturbations evolve on the time scale of the time delay $\tau_{k}$. Although these results are relevant for any systems with different and well-separated time delays, especially interesting is the application to the network of networks, where time delays within a subnetwork are shorter than the corresponding time delays between the different subnetworks. We showed, that in such a case, the units within a subnetwork can only synchronize when strong chaos is absent and the maximal LE scales either with the shorter or with the longer delay. The total synchronization of all elements involving also all subnetworks is, however, possible only when the whole LE spectrum scales with the longer delay.

\bibliography{biblio}

\end{document}